\documentclass[12pt,a4paper]{article}
\usepackage{graphicx}
\usepackage[cp1251]{inputenc}
\usepackage[english]{babel}
\usepackage{amssymb, amsfonts, amsmath}
\usepackage{amscd}
\begin{document}

\date{}

\title{Maxwell equations in homogeneous spaces with solvable groups of motions.}

\author{V. V. Obukhov}

\maketitle

\quad Institute of Scietific Research and Development,
Tomsk State Pedagogical University (TSPU),  60 Kievskaya St., Tomsk, 634041, Russia; \\ \quad

Laboratory for Theoretical Cosmology, International Center of Gravity and Cosmos, Tomsk State University of Control Systems and Radio Electronics (TUSUR), 36, Lenin Avenue, Tomsk, 634050, Russia



\quad

Keywords: Maxwell equations, Klein-Gordon-Fock equation, algebra of symmetry operators, theory of symmetry, linear partial differential equations.

\section{Introduction} A special place in mathematical physics is occupied by the problem of the exact integration of the field equations for electromagnetic and gravitational fields. The~ problem can be successful solved if the space and the electromagnetic fields possess some symmetry. Homogeneous spaces are one of the important examples of the space manifolds with symmetry. Stackel spaces are another example of such spaces. Both of these sets of spaces are applied in the theory of electromagnetism and gravitation due to the fact that, in these spaces, methods of commutative and noncommutative integration of equations of motion of single test particles can be~applied.

The methods of commutative integration is based on the use of a commutative algebra of symmetry operators (integrals of motion) that form a complete set. The~complete set includes first- and second-degree linear operators in momentum formed from complete sets of geometric objects consisting of vector and tensor Killing fields. The~method is known as the method of the complete separation of variables. The~theory of the complete separation of variables was mainly constructed in the works~\cite{1,2,3,4,5,6,7}. A~description of the theory and detailed bibliography can be found in~\cite{8,9,10} Examples of applications of the theory of complete separation of variables in the theory of gravitation can be found in the works~\cite{11,12,13,14,15,16}.
The methods of non-commutative integration is based on the use of the algebra of symmetry operators, which are linear in momenta and constructed using noncommutative Killing vector fields forming noncommutative groups of motion of spacetime $G_3$. Among~these spacetime manifolds, the homogeneous spaces are of greatest interest for the theory of gravity (see, for~example,~\cite{17,18,19,20,21,22,23,24,25,26,27}). The~theory of the noncommutative integration method and development of the theory can be found in the works~\cite{29,30,31,32,33,34}.

Thus, these two methods are essentially complementary and have similar classification problems (by solving a classification problem, we mean enumerating all metrics of the corresponding spaces that are not equivalent in terms of admissible transformations of privileged coordinate systems; likewise, all electromagnetic potentials of admissible electromagnetic fields that are not equivalent in terms of admissible gradient transformations).
Among these classification problems, the most important are the~following.

The classification of all metrics of the Stackel and homogeneous spaces in privileged coordinate systems. For~Stackel spaces, this problem was solved in the papers cited above. For~homogeneous spaces, this problem was solved in the work of Petrov (see~\cite{28}).

The classification of all (admissible) electromagnetic fields to which these methods can be applied. For~the Hamilton--Jacobi and Klein--Gordon--Fock equations, this problem is completely solved in homogeneous spaces (see~\cite{30,31,32,33}). In~Stackel spaces, it is completely solved for the Hamilton--Jacobi equation (see~\cite{8,9,10}) and partially solved for the Klein--Gordon--Fock~equation.

The classification of all vacuum and electrovacuum solutions of the Einstein equations with metrics of Stackel and homogeneous spaces in admissible electromagnetic fields. This problem is completely solved for the Stackel metric (see, for~example,~\cite{5,12,13} and bibliography in~\cite{8,9,10}). For~homogeneous spaces, this classification problem has not yet been~studied.

Thus, for~the complete solution of the problem of uniform classification, it remains to integrate the Einstein--Maxwell vacuum equations using the previously found potentials of admissible electromagnetic fields and the known metrics of homogeneous spaces in privileged (canonical) coordinate systems. This problem can also be divided into two stages. In~the first stage, all solutions of Maxwell vacuum equations for the potentials of admissible electromagnetic fields should be~found.

In the paper~\cite{35}, the first problem was decided for the case where there exist groups $G_3(II$--$VI)$ in the homogeneous spaces. The~present work is devoted to the homogeneous spaces with groups of motion $G_3(VII)$. Thus, the classification problem for solvable groups of motions will be~solved.

\section{Maxwell Equations in the Homogeneous~Spaces}
\unskip

\subsection*{Homogeneous~Spaces}

By definition, a~space--time manifold $V_4$ is a homogeneous space if a three-parameter group of motions acts on it whose transitivity hypersurface $V_3$ is endowed with the Euclidean space signature.
A semi-geodesic coordinate system $[u^i] $ is used. The~metric $V_4$ has the form:
\begin{equation}\label{1a}
ds^2 = g_{ij}du^i du^j =-{du^0}^2 + g_{\alpha \beta}du^\alpha du^\beta, \quad det|g_{\alpha\beta}|>0.
\end{equation}

 Coordinate indices of the variables of the semi-geodesic coordinate system are denoted by lower-case Latin letters: $i, j,\ldots = 0, 1 \ldots 3$.  The coordinate indices of the variables of the local coordinate system on the hypersurface  $ V_3$   are denoted by lower-case Greek letters: $\alpha, \beta, \gamma,\ldots =1, \ldots 3$.   The temporal variable is indexed by 0. Group indices and indices of a non-holonomic frame are denoted by  $a, d, c \ldots = 1, \ldots 3$.  The letters   p, q   denote the indices varying from 2 to 3. Summation is performed over repeated upper and lower indices within the index~range.

Another definition of a homogeneous space exists, according to which, the spacetime  $V_4$   is homogeneous if its subspace  $V_3$,   endowed with the Euclidean space signature, admits a set of coordinate transformations (the group $G_3$ of motions spaces $V_4$) that allow us to connect any two points in  $V_3$ (see, e.g.,~\cite{34} ). This definition directly implies that the metric tensor of the $V_3$ space can be represented as follows:
\begin{equation}\label{2a}
g_{\alpha\beta}=e^a_\alpha e^b_\beta \eta_{ab}(u^0),   \quad e^a_{\alpha, 0}=0,  \quad \eta_{ab} =\eta_{ab}(u^0).
\end{equation}
while the form
$$\omega^a= e^a_\alpha du^\alpha$$
is invariant with respect to transformations of the group $G_3$. The~vectors of the frame $e^a_\alpha$ define a non-holonomic coordinate system in $V_3$. The~dual triplet of vectors $e_a ^\alpha$
$(e_a ^\alpha e^b_\alpha =\delta_a^b, ~e_a ^\alpha e^a_\beta = \delta^\alpha_\beta) $
constructs the operators of the $G_3$ algebra group:
\begin{equation}\label{3a}
\hat{Y}_a= e_a ^\alpha \partial_a, \quad  [\hat{Y}_a,\hat{Y}_b] = C_{ab}^c \hat{Y}_c.
\end{equation}

In the following, this definition of homogeneous spaces is used. The~electromagnetic field is invariant with respect to transformations of the group acting in the space. It has the form:
\begin{equation}\label{4a}
  A_i = l^a_i\alpha_a \quad \alpha_a, = \alpha_a(u^0).
\end{equation}

\section {Maxwell Equations}
We consider the Maxwell equations with zero sources for electromagnetic potential~\eqref{4a}:
\begin{equation}\label{1b}
\frac{1}{\sqrt{-g}}(\sqrt{-g}F^{ij})_{,j} = 0.
\end{equation}
\begin{equation}\label{2b}
 \frac{1}{\sqrt{-g}}(\sqrt{-g}g^{\alpha\beta}F_{0\beta})_{,\alpha} = \frac{1}{l}(l l^{\alpha}_{a}\eta^{ab}\dot{\alpha}_{b})_{,\alpha} = (l^\alpha_{a,\alpha} + \frac{l_{|a}}{l})\frac{\beta^a}{\eta}\quad  (\beta^a = \eta^{ab}\eta\dot{\alpha}_b).
\end{equation}
Notation used:
$$
f_{|a} = l^\alpha_a f_{,\alpha}, \quad  g=-\det|g_{\alpha\beta}|=-(\eta l)^2, \quad (\eta^2 = \det|\eta_{\alpha\beta}|,  \quad l=\det|l_{\alpha}^a|).
$$
The dots denote the time derivatives. Then, we have the first equation in the form:
\begin{equation}\label{3b}
(l^\alpha_{a,\alpha} + l_{|a})\beta^a =0.
\end{equation}

{ If $i=\alpha$, from Equation \eqref{1b}, it {follows} that}: 
\begin{equation}\label{4b}
\frac{1}{\eta}(\eta g^{\alpha\beta}F_{0\beta})_{,0} = \frac{1}{l}(lg^{\nu\beta}g^{\alpha\gamma}F_{\beta \gamma})_{,\nu} \Rightarrow \frac{1}{\eta}(\eta \eta^{ab}l^\alpha_a \dot{\alpha}_{b})_{,0}=\frac{1}{l}(l l_a^\nu l^\beta_b \eta^{ab} l_{\tilde{a}}^{\alpha} l_{\tilde{b}}^{\gamma} \eta^{\tilde{a}\tilde{b}}F_{\beta\gamma})_{,\nu}\Rightarrow
\end{equation}
\begin{equation}\label{5b}
\frac{l^\alpha_a}{\eta}\dot{\beta}^a = \frac{1}{l}(l l^\beta_b l_{\tilde{a}}^{\alpha} l_{\tilde{b}}^{\gamma}F_{\beta\gamma})_{|a}\eta^{ab}\eta^{\tilde{a}\tilde{b}}.
\end{equation}
 $F_{\alpha\beta}$   can be found using  the relations \eqref{2a}--\eqref{4a}:
\begin{equation}\label{6b}
F_{\alpha\beta} = (l^a_{\beta,\alpha}-l^a_{\beta,\alpha})\alpha_a = l^c_\beta l^\gamma_c l^d_\alpha l^\nu_d(l^a_{\gamma,\nu}-l^a_{\nu,\gamma})\alpha_a=l^b_\beta l^a_\alpha l^c_\gamma  (l^\gamma_{a|b}-l^\gamma_{b|a})\alpha_c = l^b_\beta l^a_\alpha C^c_{ba}\alpha_c  \quad \Rightarrow
\end{equation}
\begin{equation}\label{7b}
(lF^{\alpha\beta})_{,\beta} = \eta^{ab}\eta^{\tilde{a}\tilde{b}}C^d_{\tilde{b}b}\alpha_d ((ll^\alpha_a)_{|\tilde{a}} + ll^\alpha_al^\gamma_{\tilde{a},\gamma}).
\end{equation}
Structural constants of a  group $G_3$ can be represent in the form:
\begin{equation}\label{8b}
C^c_{ab} = C^c_{12}\varepsilon^{12}_{\tilde{a}\tilde{b}} + C^c_{p3}\varepsilon^{p3}_{\tilde{a}\tilde{b}},
\end{equation}
where
$$
\varepsilon^{AB}_{ab} = \delta^A_a\delta^B_b - \delta^A_b\delta^B_a.
$$

From the relations:
\begin{equation}\label{9b}
 (\varepsilon^{AB}_{\tilde{a}\tilde{b}}\eta^{a\tilde{a}}\eta^{b\tilde{b}}) =(\eta^{aA}\eta^{bB}-\eta^{aB}\eta^{bA}),
\end{equation}
it follows that:
$$
\eta^2\varepsilon^{12}_{cd}\eta^{ac}\eta^{bd} =(\eta_{33}\varepsilon^{ab}_{12} + \eta_{23}\varepsilon^{ab}_{31} +\eta_{13}\varepsilon^{ab}_{23}),
$$
$$
\eta^2\varepsilon^{31}_{cd}\eta^{ac}\eta^{bd} =(\eta_{22}\varepsilon^{ab}_{31}+ \eta_{23}\varepsilon^{ab}_{12}+\eta_{12}\varepsilon^{ab}_{23}),
$$
$$
\eta^2\varepsilon^{23}_{cd}\eta^{ac}\eta^{bd} =(\eta_{13}\varepsilon^{ab}_{12} + \eta_{12}\varepsilon^{ab}_{31}+\eta_{11}\varepsilon^{ab}_{23}).
$$
Equations \eqref{1b} take the form:
\begin{equation}\label{10b}
\eta\dot{\beta}^a = \delta^a_1(\gamma_1C^1_{32} - \gamma_2(C^1_{31} +\omega_3) + \gamma_3(C^1_{21} +\omega_2)) + \delta^a_2(\gamma_1(C^2_{32} +\omega_3) +
\end{equation}
$$
\gamma_2C^2_{13} - \gamma_3(C^2_{12} +\omega_1)) +\delta^a_3(-\gamma_1(C^3_{23} +\omega_2) + \gamma_2(C^3_{13} +\omega_1) + \gamma_3C^3_{21}),
$$
\begin{equation}\label{11b}
 \eta_{ab}\beta^b = \eta \dot{\alpha}_a,
\end{equation}
\begin{equation}\label{12b}
\omega_a \beta^a =0,\quad \omega_a=l^\alpha_{a,\alpha} + l_{|a}/l,
\end{equation}
where
$$
\gamma_1=\sigma_1\eta_{11}+\sigma_2\eta_{12}+\sigma_3\eta_{13},\quad  \gamma_2=\sigma_1\eta_{12}+\sigma_2\eta_{22}+\sigma_3\eta_{23},
$$
$$
\gamma_1=\sigma_1\eta_{13}+\sigma_2\eta_{23}+\sigma_3\eta_{33},\quad
\sigma_1 = C^a_{23}\alpha_a, \quad \sigma_2 = C^a_{31}\alpha_a, \quad  \sigma_3 = C^a_{12}\alpha_a.
$$
Let us find sets of the Maxwell Equations \eqref{10b}--\eqref{12b} for all solvable~groups.

\subsection*{ Groups $  G_3(I$--$VII)$}
The components of the metric tensor and structural constants  $C^c_{ab}$   were found by Petrov (see~\cite{28}). The~components of the vector   $l^\alpha_a$   were found in our work~\cite{35}:
\begin{equation}\label{14b}
e^\alpha_a = \delta^1_a \delta^\alpha_1\exp(-ku^3)+\delta^2_a (-\delta^\alpha_1\varepsilon u^3 \exp(-ku^3) + \delta^\alpha_2 \exp(-nu^3)) + \delta^\alpha_3\delta^3_a,
\end{equation}
$$
e^a_\alpha = \delta_1^a \delta_\alpha^1\exp(ku^3)+\delta^2_a (\delta^\alpha_1\varepsilon u^3 \exp nu^3 + \delta^\alpha_2 \exp nu^3)) + \delta_\alpha^3 \delta^3_a,
$$
\begin{equation}\label{15b}
C^c_{ab}=k\delta^c_1\varepsilon^{13}_{ab}+(\varepsilon\delta^c_1 +n\delta^c_2)\varepsilon^{23}_{ab}.
\end{equation}
Let us consider Maxwell Equations \eqref{10b}--\eqref{12b}.

\quad

{\bf I.} For the groups G(I-VI), the equations can be presented in the form:

{{(1)}}  For the group\quad $ G_1(I) (k=n=\varepsilon=0):$
$$
\dot{\beta}^a = 0,  \dot{\alpha}_a = \frac{1}{\eta}\eta_{ab}\beta^b \Rightarrow
$$

Solution of the Maxwell Equations \eqref{10b}--\eqref{12b} has the form:
\begin{equation}\label{16b}
\beta^a = const, \quad \alpha_a = \beta^b\int\frac{1}{\eta}\eta_{ab}du^0;
\end{equation}

{ {(2)}}  For the group $  G_1(II) \quad (k=n=0,  \quad \varepsilon=1)$:
\begin{equation}\label{17b}
\dot{\beta}^a = -\delta^a_1\alpha_1 \eta_{11}, \quad \dot{\alpha}_a = \frac{1}{\eta}\eta_{ab}\beta^b;
\end{equation}

{{(3)}}  For the group $ G_1(III) \quad (k=1,  \quad n=\varepsilon=0)$:
\begin{equation}\label{18b}
\dot{\beta}^a = -\delta^a_1 \alpha_1\eta_{22}, \quad \beta^3 = 0,  \dot{\alpha}_a = \frac{1}{\eta}\eta_{ab}\beta^b;
\end{equation}

{ {(4)}}  For the group $ G_1(IV)\quad  (k= n=\varepsilon=1)$:
\begin{equation}\label{19b}
\dot{\beta}^a = -\delta^a_1 ((\alpha_1 + \alpha_2)\eta_{11} + \alpha_2eta_{12} -\alpha_1 \eta_{22}) +\delta^a_2 ((\alpha_1 + \alpha_2)\eta_{11}-\alpha_1\eta_{12});
\end{equation}
$$
\beta^3 = 0,
 \dot{\alpha}_a = \frac{1}{\eta}\eta_{ab}\beta^b;
$$

{{(5)}}  For the group $  G_1(V) \quad (k= n=1, \quad \varepsilon=0$):
\begin{equation}\label{20b}
\dot{\beta}^a = \delta^a_1 (-\alpha_2\eta_{12} + \alpha_1\eta_{22})  +\delta^a_2 (\alpha_1 \eta_{12}-\alpha_2\eta_{11}),\quad \beta^3 = 0,\quad
 \dot{\alpha}_a = \frac{1}{\eta}\eta_{ab}\beta^b;
\end{equation}

{{(6)}}  For the group $G_1(VI)\quad (k=1, \quad  n=2,\quad \varepsilon=0)$:
\begin{equation}\label{21b}
\dot{\beta}^a = -\delta^a_1 (2\alpha_2\eta_{12} - \alpha_1\eta_{22})  +\delta^a_2 (2\alpha_2\eta_{11}-\alpha_1 \eta_{12}),\quad \beta^3 = 0,\quad
 \dot{\alpha}_a = \frac{1}{\eta}\eta_{ab}\beta^b.
\end{equation}

{}Equations \eqref{17b}, \eqref{21b} were integrated into our work~\cite{35}. In~the present paper, the~solutions for the group $G(VII)$ were~found.

{\bf {II.}}   Group $G(VII)$.

When obtaining  the Maxwell equations for the groups  $G_3(I$--$VI)$,  the components of vector fields  $ l_a^\alpha$   could be constructed directly from the components of the metric tensor (see~\cite{35}). For~the group  $G(VII)$,   this cannot be performed. Therefore, the vectors  $l^\alpha_a$   must be found directly from the conditions \eqref{2a}. Consider these conditions for the structural constants of the group $G_3(VII)$:
$$C^a_{23}=-\delta^a_1 + 2\delta^a_2 \cos\alpha,  \quad C^2_{13} = 1, \quad \alpha =const.
$$

 By coordinate transformation of the form  $\tilde{u}^\alpha = \tilde{u}^\alpha(u^\beta)$ the vector field  $l^\alpha_3$ can be diagonalized:
$$l^\alpha_3 = \delta^\alpha_3.$$

{}  From the commutation relations, it follows that: 
\begin{equation}\label{22b}
X_{1,3}=-X_2;  \quad X_{2,3}=X_1-2X_2\cos\alpha \Rightarrow l_2^\alpha = -l^\alpha_{1,3},  \quad  l_{2,33}^\alpha + 2l^\alpha_{1,3}\cos\alpha +l^\alpha_1 = 0.
\end{equation}

{} Solution of the Equation \eqref{22b} has the form:
$$
l^\alpha_1 = \exp{(-q_3)}(a^\alpha_1(u^p)\sin{p_3} + b^\alpha_1(u^p)\cos{p_3}),
$$
$$
l^\alpha_2 = -\exp{(-q_3)}(a^\alpha_2(u^p)\sin{(p_3-\alpha)} + b^\alpha_2(u^p)\cos{(p_3-\alpha)}),
$$
where  $p, q = 1,2,   q_3=u^3\cos\alpha,   p_3=u^3\sin\alpha$.
Since the operators  $X_p$  commute, the~vectors $  a^p_q,    a^p_q$   can be simultaneously diagonalized by coordinate transformations of the form  $\tilde{u}^p=\tilde{u}^p(u^q)$:
$$
a^p_q = \delta^p_q,  \quad b^p_q = \delta^p_q,
$$

From the commutation relations it follows that: $a^p_3 =0,   b^p_3 =0$.

Thus, the~vectors of the frame of the homogeneous space of type $VII$ according to Bianchi can be represented in the form:
\begin{equation}\label{23b}
l^\alpha_1 = \exp{(-q_3)}(\delta^\alpha_1\sin{p_3} + \delta^\alpha_2\cos{p_3}),
\end{equation}
$$l^\alpha_2 = \exp{(-q_3)}(\delta^\alpha_1\sin{(p_3-\alpha)} + \delta^\alpha_2\cos{(p_3-\alpha)}), \quad  l^\alpha_3 = \delta^\alpha_3.
$$

The Maxwell Equations will take the form:

\begin{equation}\label{24b}
 \eta\dot{\beta}_a=\delta^a_1(\gamma_1 -2\gamma_2\cos\alpha))+\delta^a_2\gamma_2, \quad  \Rightarrow  \quad \gamma_2=\eta\dot{\beta}_2,  \gamma_1=\eta(\dot{\beta}_1 + 2\dot{\beta}_2\cos\alpha).
\end{equation}
The system of Maxwell's equations can be represented in the form:
\begin{equation}\label{25b}
\sigma\eta_{11}-\alpha_2\eta_{12}=\gamma_1,  \quad \sigma\eta_{12}-\alpha_2\eta_{22}=\gamma_2  (\sigma=2\alpha_2\cos{\alpha}-\alpha_1);
\end{equation}
\begin{equation}\label{26b}
\beta_1\eta_{11}+\beta_2\eta_{12}=\eta\dot{\alpha}_1,  \quad \beta_1\eta_{12}+\beta_2\eta_{22}=\eta\dot{\alpha}_2,  \beta_3=0;
\end{equation}
\begin{equation}\label{27b}
\eta\dot{\alpha}_3=\beta_1\eta_{13}+\beta_2\eta_{23} \quad  \Rightarrow \quad  \alpha_3=\int\frac{\beta_1\eta_{13}+\beta_2\eta_{23}}{\eta}du_0. \end{equation}

From Equations \eqref{25b} and \eqref{26b}, it follows that:
\begin{equation}\label{28b}
\eta_{11}(\alpha_2 \dot{\alpha}_2 -\sigma \dot{\alpha}_1)(\alpha_2\beta_1 + \sigma\beta_2) = \gamma_1\beta_2(\alpha_2 \dot{\alpha}_2 -\sigma \dot{\alpha}_1) - \alpha_2\dot{\alpha}_2(\beta_1\gamma_1 + \beta_2\gamma_2).
\end{equation}
\begin{equation}\label{29b}
\alpha_1\dot{\alpha}_2(\eta(\alpha_2 \dot{\alpha}_2 -\sigma \dot{\alpha}_1) +\beta_1\gamma_1 + \beta_2\gamma_2 ) =0.
\end{equation}
When solving the system of equations \eqref{28b}, \eqref{29b}, the variants that need to be considered are:

{\bf {A}}. $\alpha_2 \ne 0$.
 From the system of Equation \eqref{26b}, it follows:
\begin{equation}\label{30b}
 \eta_{11}(\alpha_2\beta_1 + \sigma\beta_2) = \eta(\dot{\alpha}_1\alpha_2 + \dot{\beta}_1\beta_2 ),  \quad
\eta_{12} = \frac{1}{\alpha_2}(\sigma_1 \eta_{11} - \eta\tilde{\beta}_1), \quad  \eta_{22} = \frac{1}{\alpha_2^2}(\sigma_1^2 \eta_{11} - \eta (\sigma_1\tilde{\beta}_1 + \alpha_2 \dot{\beta_2})).
\end{equation}
When solving the set of Equations \eqref{28b} and \eqref{30b}, the following variants must  be consider:

\quad

{\bf {1.}} $(\alpha_2 \dot{\alpha}_2 -\sigma \dot{\alpha}_1)\ne 0 \quad \Rightarrow \quad
\eta_{11} =\eta\frac{\dot{\alpha}_1\alpha_2 + \dot{\beta}_1\beta_2}{\alpha_2\beta_1 + \sigma\beta_2}.
$   We consider Equation \eqref{29b}:
Let us use the following notations:
$$
\alpha_1=\sqrt{\rho}\sin{(\omega/2)},  \quad \alpha_2=\sqrt{\rho}\cos{(\omega/2)},  \quad
\Omega=(2\beta_2\dot{\beta}_1\cos\alpha + \beta_1\dot{\beta}_1 + \beta_2\dot{\beta}_2), \quad \omega = \omega(u^0),$$
{ {(1)}}  Let  $\alpha_1 \ne 0 $.   Then the equation \eqref{31b} can be reduced to the form:
\begin{equation}\label{31b}
2\alpha_2\dot{\alpha}_1\cos\alpha - \alpha_1\dot{\alpha}_1 - \alpha_2\dot{\alpha}_2 = 2\beta_2\dot{\beta}_1\cos\alpha + \beta_1\dot{\beta}_1 + \beta_2\dot{\beta}_2.
\end{equation}

 {{(a)}} $\dot{\omega} \ne 0.$   In this case, we take the function  $\omega$   as a new time variable and denote by the point the derivative on this variable. The~functions  $\beta_p,  \rho $  depend on $\omega$.
Then the equation \eqref{31b} can be reduced to the form:
\begin{equation}\label{32b}
\dot{\rho}(\cos\alpha \sin\omega -1)+\cos\alpha(1+\cos\omega)\rho = 2\Omega.
\end{equation}

The function  $\rho$ can be represented in the form: 
$\rho = \mathfrak{R}(\omega)\tau(\omega)$,
where
$$
\mathfrak{R}=\int\frac{\cos\alpha(1+\cos\omega)}{1-\cos\alpha \sin\omega}d\omega,
$$

The function $\tau$ has the form:
$$
\tau=(c + 2\int\frac{\Omega}{\mathfrak{R}(1-\cos\alpha \sin\omega)}d\omega),
$$

{(b)} $\omega=a =const  \Rightarrow  \rho=(c - 2\int\frac{\Omega}{(1-\cos\alpha \sin\omega)}du^0), $

\quad

{{\bf 2.}  $\alpha_1 = 0, ~~\eta_{11}
=\eta\frac{\beta^2 \dot{\beta}}{\alpha_2}, ~~ \eta_{12}=-\eta\frac{\beta^1 \dot{\beta}}{\alpha_2}, ~~ \eta_{22}=-\eta\frac{\dot{\beta}^2 + 2\cos{\alpha}\beta^1\dot{\beta}}{\alpha_2}, ~~\beta = \ln{(\beta^1 + 2 \cos{\alpha}\beta^2)}$}

The final solutions are represented in  {\bf Solutions}.

{\bf 3.}  $\alpha_2\beta_1 + \sigma\beta_2 =0,  \eta_{11}$   is an arbitrary\quad function   of  $u^0$. In~this case, there are two variants to consider:

{\bf (a)} $ \alpha_1 =0 \Rightarrow$   function  $\eta_{pq}$   can be found from \eqref{30b}.

{(b)} $ \alpha_1 \ne 0 \Rightarrow \dot{\alpha}_1\alpha_2 +\dot{\beta}\beta_2 = \dot{\alpha}_2\alpha_2 +\dot{\beta}_2\beta_2 =0  \Rightarrow  \alpha_2\dot{\alpha}_2+\beta_2\dot{\beta}_2 =0  (\beta = 2\beta_2\cos\alpha + \beta_1).$

From the last equation it follows that:
$$
\alpha_2 = c\sin \omega  \beta_2 = c\cos \omega.
$$
\qquad

{\bf B.}  $\alpha_2 = 0$.
In this case, from the set of Equations \eqref{25b} and \eqref{26b},  it follows that:
$$
\alpha_1\dot{\alpha_1} + \beta_1\dot{\beta_1} + 2\cos\alpha\beta_1\dot{\beta_2}=0, \quad  \alpha_1 = \sqrt{c-(\beta^1)^2} - 4 \cos\alpha\int\beta_1\dot{\beta}_2 du^0.
$$
The functions  $\eta_{ab}$  are determined from Equations \eqref{25b} and \eqref{26b}. The~results are given in the {\bf Solutions}.

\section{Solutions}

In this section, all solutions of Maxwell's vacuum equations for homogeneous Bianchi type VII spaces and electromagnetic fields invariant with respect to the groups of motions  $G_3(VII)$   are given. For~all solutions, the functions  $\alpha_3$   and  $\eta_{33}$   have the form:
$$ \alpha_3 = \int{(\eta_{13}\beta^1}+ {\eta_{23}\beta^2)}du^0, \quad
\eta_{33}=\frac{\eta^2 -2\eta_{12}\eta_{13}\eta_{23}+\eta_{11}\eta_{23}^2 + \eta_{22}\eta_{13}^2}{\eta_{11}\eta_{22} - \eta_{13}^2}.
$$

 Other functions that specify solutions are shown~below.

{\bf 4.1} $\alpha_2 \ne 0.$

{The functions}
  $\eta_{12}$,  $\eta_{22}$   have the form:

$$
\eta_{12} = \frac{1}{\alpha_2}(\sigma_1 \eta_{11} - \eta\dot{\beta}),  \quad \eta_{22} = \frac{1}{\alpha_2^2}(\sigma_1^2 \eta_{11} - \eta (\sigma_1\dot{\beta}^1 + \alpha_2 \dot{\beta})), \quad  \sigma_1 = 2\alpha_2\cos{}\alpha -\alpha_1,  \quad \beta = 2\beta^2\cos\alpha + \beta^1.
$$

{(1)} $\beta^1\alpha_2+\beta_2\sigma_1 \ne 0,
\eta_{11}= \eta\frac{\dot{\alpha}_1\alpha_2 + \dot{\beta}^2\dot{\beta}}{\beta^1\alpha_2+\beta_2\sigma_1},
\Omega = (\beta^1\dot{\beta}^1 + \beta^2\dot{\beta}^2) + 2\beta^2\dot{\beta}^1\cos{\alpha}.
$

{(a)} $ \alpha_1 = \sqrt{\rho}\sin c  \quad \alpha_1 = \sqrt{\rho}\cos c,  \quad  \rho = \int\frac{2\Omega du^0}{\cos{\alpha}\sin{c}-1}. $

{(b)}  $\alpha_1=\sqrt{\rho}\sin{\frac{\omega}{2}},  \quad \alpha_2=\sqrt{\rho}\cos{\frac{\omega}{2}}, \quad \omega =\omega(u^0), \quad \beta^p=\beta^p(\omega),   \quad  \dot{\beta}^p=\partial\beta^p/\partial\omega,
$
$$
\rho  = \frac{\mathfrak{R}}{1-\cos\alpha \sin\omega} (c -2\int\frac{  \Omega (1-\cos\alpha \sin\omega)d\omega }{\mathfrak{R}}),
 \mathfrak{R}= \exp{\int\frac{\cos{\alpha} d\omega}{1-\cos{\alpha}\sin{\omega}}}
$$

{(c)}
$\alpha_1 = 0, ~\eta_{11}
=\eta\frac{\beta^2 \tilde{\beta}}{\alpha_2},~
 \eta_{12}=-\eta\frac{\beta^1 \tilde{\beta}}{\alpha_2}, ~\eta_{22}=-\eta\frac{\dot{\beta}^2 + 2\cos{\alpha}\beta^1\tilde{\beta}}{\alpha_2} \tilde{\beta} = (\ln(\beta^1 + 2 \cos{\alpha}\beta^2))_{,0}.$

{(2)} $ \eta_{11}$   is an arbitrary\quad function   of  $u^0$.

 {(a)} $ \alpha_1 =0,~\eta_{12}=2\eta_{11}\cos\alpha -\eta,
\eta_{22}=4\eta_{11}\cos^2\alpha - \eta\frac{2\dot{\beta}\cos\alpha +\dot{\beta}}{\alpha_2}.
$

{(b)}
$
\alpha_1 = ac\sin\omega, ~\alpha_2 = c\sin\omega,~\beta_2 =c\cos\omega, ~\beta_1 = c(a-2\cos\alpha)\cos\omega. ~c, a=const
$
$$
\eta_{12}=(2\cos\alpha -a)\eta_{11} + a\eta,  ~\eta_{22}=(2\cos\alpha -a)^2\eta_{11} + \eta(a(2\cos\alpha -a) + 1).
$$

{\bf 4.2} $\alpha_2=0.$

{1.} $  \alpha_1 = \sqrt{c-(\beta^1)^2 - 4 \cos\alpha\int\beta^1\dot{\beta}^2 du^0.} ~
\eta_{11} = -\eta\frac{ (2\cos \alpha \dot{\beta}_2 +\dot{\beta}_1) }{\alpha_1}, ~ \eta_{12} = -\eta\frac{\dot{\beta}_2}{\alpha_1},  ~
 \eta_{22} = \eta\frac{\dot{\beta}_2\beta_1}{\beta_2\alpha_1}.$

{2.} $ \beta_2 = 0,  \alpha_1=c\sin\omega, \beta_1=c\cos\omega. \eta_{22},  \omega$   are arbitrary\quad functions   of  $u^0$.
$$
\eta_{12}=0,  \quad \eta_{11}=-\eta\dot{\omega}, \quad  \eta_{13}=\eta\frac{\dot{\alpha}_3}{\beta_1}.
$$

All functions included in these expressions that are not additionally described (for example, $\eta,  \eta_{p3}$,   and so on) are arbitrary functions of  $u^0.$

\section{Conclusions}
In the paper, the classification of solutions of vacuum Maxwell equations for the case where the electromagnetic fields and the metrics of homogeneous spaces are invariant with respect to solvable groups of motions was completed (for the groups $G_3(I$--$VI)$, classification was carried out in the paper~\cite{35}).  Since this classification was carried out in the canonical frame  \eqref{2a}, it allows one to proceed with the classification of exact solutions of the vacuum Einstein--Maxwell equations for the found fields. This will be of interest for the study of the early stages of the evolution of the~Universe.

\quad

FUNDING: The work is supported by Russian Science Foundation, project number N 23-21-00275.

INSTITUTIONAL REVIEW BOARD STATEMENT: Not applicable.

INFORMED CONSENT STATEMENT: Not applicable.

DATA AVAILABILITY STATEMENT: The data that support the findings of this study are available within the article.

ACKNOWLEDGMENTS: The work has no financial support
Conflicts of Interest: The author declares no conflict of interest.

\end{document}